# The gravitational braking of captured moons around ringed planets

George Bell

The University of Lincoln

Brayford Pool, Lincoln, LN6 7TS, UK

25189120@students.lincoln.ac.uk

## ABSTRACT

Irregular moons are a class of satellite found orbiting all of the Solar System's giant planets: as their orbits don't match those of their planets, they are theorised to have formed elsewhere in the Solar System and were subsequently captured into their observed orbits.

Missions such as Cassini have contributed significant empirical data on irregular moons in the present day but this paper aims to develop our currently limited theoretical understanding of their origins and capture as it presents one of the first projects to connect moon capture with another feature common to all giant planets: ring systems.

As a captured body gravitationally brakes around a ringed planet, it transfers orbital energy to the planetary system, a process which has been seen to leave distinctive signatures on the rings which may be used to constrain key parameters of this interaction, including the trajectory and timing. This paper presents a project which applies this technique to constrain scenarios for moon capture through conducting a series of computational simulations using the Python version of the astrophysical code REBOUND modelling the capture of the large irregular moon Phoebe by the planet Saturn and Phoebe's effect on Saturn's ring system.

By helping to constrain scenarios for moon capture, this research will further our understanding of the moon systems of the giant planets while simulating the effects of a moon's interaction with a ring system will offer insight into the formation and evolution of planetary rings, whether within our own Solar System or orbiting exoplanets.

## INTRODUCTION: IRREGULAR MOONS

Captured or irregular moons are a class of satellite found orbiting all four of the solar system's giant planets. They are distinguished from the regular satellites by their orbital parameters, including orbiting at greater distances from the planet, elongated orbits and orbital planes tilted at an angle to the plane of regular satellite orbits [1]. Most irregular satellites orbit in the opposite (retrograde) direction to regular satellites and planetary rings and hence could not have been formed in situ by the same processes [1].

The differences between these satellite populations has led to the prevalent theory that irregular satellites did not form with the rings and regular moons but were instead captured into orbits around the giant planets [1], a process likely concurrent with the Nice model instability in the early Solar system [2].

### Phoebe and Saturn

This paper will focus on the capture of the retrograde irregular moon Phoebe by Saturn. Phoebe is Saturn's most massive irregular moon, containing ~98% of the mass of the outer Saturn system [1] and the third largest in the Solar system. Phoebe's orbital and physical parameters, unlike those of most irregular satellites, have been well-characterised through extensive Cassini observations [1]; this substantial empirical data makes Phoebe suitable for refining theoretical models. Irregular moons of a similar size to Phoebe are found around all of the giant planets [1] while smaller moons are considered to be fragments of larger shattered moons, indicating that Phoebe is representative of the captured moon population. However, the details of Phoebe's origins and capture have not been explored in the context of the current Nice model despite its suitability as a constraint on the conditions which resulted in the observed irregular moon population.

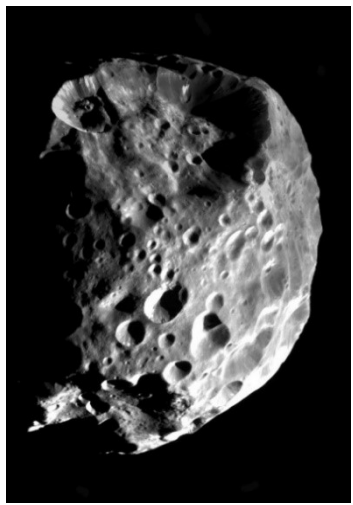

**Figure 1: "The Face of Phoebe" by Cassini [3]**

### Irregular Moons and Planetary Rings

Although ring systems exist around all of the Solar system's giant planets in the current day and are expected to have been present during the Nice model instability [4], prior work on the dynamic capture of moons has only considered the effects of this process on the planet-satellite system. Moon-ring interactions have been shown to be key to many ring features including gaps, spiral waves and corrugations while satellite observations have demonstrated the impact of captured bodies on two planetary ring systems [5] [6], reinforcing the significance of studying this missing component of moon capture. As Saturn's prominent ring system has proved suitable for showing the effects of capture [5], this provides a further motivation for a closer study of Phoebe's history.

### Organization of this Paper

Section two of this paper will set out the theoretical basis of this research, section three will consider a realistic methodology for the capture of a moon and its subsequent interaction with planetary rings and section 4 will summarize the findings to date.

## SECTION 2: THEORETICAL BASIS

### The Formation of the Solar System

Current models of planetary formation and observations of exoplanetary systems indicate the broadly spaced and low eccentricity orbits of the Solar system's four giant planets are atypical: the currently accepted explanation for this observation is the Nice model [7]. The Nice model proposes that the giant planets in our Solar system formed in a compact space where pairs of planets entered into orbital resonances with each other; this chain of resonances then became dynamically unstable when Jupiter and Saturn crossed their 2:1 resonance, causing Uranus and Neptune to be scattered outwards into the Kuiper Belt; the giant planets then migrated to their observed orbits through interaction with the Kuiper Belt [7].

More recent versions of the Nice model have included further refinements to the basic model as the violence of the instability frequently led to unsatisfactory results such as the ejection of an ice giant [7]. Assuming that an additional ice giant of similar mass to Uranus and Neptune was formed and present but was subsequently ejected during the instability (a "5GP" model) improves the probability that the final planetary orbits will resemble those observed although this highest success probability remains at a low value of 5% [7].

The approach of this paper has been informed by a recently-presented consensus 5GP Nice model, whose success criteria require the simulation to finish with four giant planets in orbits resembling those observed, that the eccentricities of Jupiter and Saturn are suitably excited and that the period ratio of Jupiter and Saturn does not exceed 2.8 so as to be consistent with the observed effects of this value on the asteroid belt [8]. This implementation is concerned with the accurate reproduction of Jupiter and Saturn's orbits and is hence less precise for constraining the gap between the outermost ice giant and the Kuiper belt [8]; however, this minor limitation is acceptable given this paper's focus on Saturn.

A further advantage of this consensus model is its consistency with a range of instability timings. The instability was previously assumed to have been the cause of the Late Heavy Bombardment (LHB) of the lunar surface ~4 Gyr ago [7]. However, this timing has since been challenged, both through reinterpretation of the data which led to the proposal of the LHB and as late instabilities pose difficulties for the formation and survival of the terrestrial planets and asteroid belt [8]. These issues may be resolved through an early instability occurring alongside the formation of the terrestrial planets within the first 100 Myr after the Solar system's formation, a scenario with which the consensus Nice model is also consistent [8].

### Irregular Satellite Capture

In all forms, the Nice model indicates that close interactions between the giant planets and Kuiper Belt objects (KBOs) were frequent and likely during the instability. Numerous mechanisms have been proposed whereby these KBOs may be captured as moons: suggestions such as collisional capture or the effects of gas drag have lower probability than would be required to adequately explain the irregular moon population hence three-body encounters are favoured [2].

This paper follows a three-body encounter capture mechanism used previously to explain Neptune's retrograde moon Triton. Here, Triton began as one component of a binary planetesimal which passed close enough to Neptune during the instability for the Hill sphere of the binary (the region where the mutual attraction of the binary exceeds the perturbing force from the larger planet) to equal the separation between the components, disrupting the binary [2]. Triton was captured into an eccentric Neptune orbit (while its partner was ejected from the Neptune system) which circularized under tidal forces to its observed orbit [2]. The resulting probability of Neptune capturing a Triton mass satellite is 0.7%, consistent with an estimate of 1000 binary planetesimals with a Triton mass component in the early Kuiper belt [2]. The tidal disruption distance (at which the binary would become ionized) is given by the below equation [2]:

$$\frac{r_{td}}{R_P} = \left(\frac{a_B}{R_1}\right)\left[\left(\frac{3\rho_p}{\rho_1}\right)\left(\frac{m_1}{m_1+m_2}\right)\right]^{\frac{1}{3}} \qquad (1)$$

Where $r_{td}$ = tidal disruption distance, $R_p$ = planetary radius, $a_B$ = semi-major axis of the binary, $R_1$ = radius of the binary's primary component, $\rho_p$ = planetary density, $\rho_1$ = density of the binary's primary component and $m_1$ and $m_2$ = respectively the masses of the primary and secondary components of the binary.

This paper applies this three-body capture to the Solar system's other prominent irregular moon, Phoebe, in the context of a newer version of the Nice model than was used in the Triton example. Existing work on Phoebe's capture in the scientific literature uses a collisional capture scenario and hence gives a very low probability [9] at odds with the presence of satellites comparable to Phoebe at all of the Solar system's giant planets [1] while the orbital and rotational data used has been superseded by more recent Cassini observations [1].

Substituting suitable values for Phoebe and Saturn into Equation 1 and, as with the Triton example, assuming that Phoebe is the primary component of a binary with a circular orbit of semi-major axis 37,146 km (chosen to be consistent with the widest known KBO binaries) [2] the Phoebe binary would tidally disrupt at ~299 Saturn radii. Compared to Triton, Phoebe experiences lower tidal forces: Phoebe's orbit has undergone much less tidal evolution and is both eccentric and distant [1].

Modelling on various astrophysical scales has found retrograde orbits to be more stable in the long term and stable for greater semi-major axes than prograde orbits [10] including at distant orbits around giant planets where orbits are strongly influenced by solar perturbations such as the Kozai effect [2].

### The Saturn System

As irregular satellite capture is thought to take place in the early Solar System [1] it is essential to determine the nature of the primordial Saturn system with which Phoebe would have interacted after capture, in particular the distribution and number of moons and the nature and presence of the ring system.

Findings from the latter stages of the Cassini mission inform this paper's approach to the latter subject. For most of Cassini's mission the probe orbited outside of Saturn's rings but the eccentric and inclined "Grand Finale" orbits in 2017 took it between the planet and the rings for the first time [11]. In these orbits the gravitational force on the spacecraft from the planet acted in the opposite direction to that from the ring system and the two forces could be calculated from the Doppler effect they induced on a microwave link between Cassini and Earth, which enabled the calculation of the current mass of the rings as $0.41 \pm 0.13$ that of Saturn's moon Mimas (which has a mass of $3.75 \times 10^{19}$ kg) [11]. This figure is much lower than the 0.75 Mimas mass figure calculated from the Voyager mission, challenging the interpretation of the rings as ancient both because rings of this low mass may not be able to withstand bombardment and as it was assumed that rings formed of "pure water ice" would darken over time due to interplanetary dust flux [11]. It was instead proposed that ring formation took place in the last 100 Myr, perhaps through the tidal disruption of a captured interplanetary body, although there was no precise explanation why this happened so recently [11].

This interpretation was quickly disputed as the apparent age of the rings from exposure and structure may be much less than the true age [4]. The measured mass could be attained by the evolution of an ancient Saturnian ring of any initial mass as the ring mass is proportional to the speed of ring spread beyond its inner and outer boundaries (which remain the same throughout Saturn's history), the process by which ring material is lost [4]. In addition, the current bombardment rate may not be typical as this parameter varies over time and many key details remain unknown while the brightness and exposure of the rings may be explained by treating the rings not as an isolated system but as interacting with their environment [4]. The presence of silicate grains and organic molecules in Saturn's atmosphere indicated that the rings may be preferentially losing dark and dusty material, to the extent that the prior argument may be reversed such that the rings formed with a higher silicate component and became brighter over time [4]. Ancient rings not only provide a better motivation for the observed rings than the reliance on a recent chance encounter; they can also account for the formation and distribution of many of Saturn's regular satellites.

The properties of Saturn's small and "mid-sized" regular moons orbiting between the rings and its largest moon Titan would be unusual if they formed simultaneously: these moons have low densities but also some degree of differentiation between icy surfaces and rocky interiors while their age (as derived from cratering frequency) is difficult to reconcile with that of the outer regular moon Iapetus [12]. The proposed solution is that these moons did not form at the same time, rather these moons formed from the rings which initially contained ice and silicates when this material spread beyond the ring's outer edge [12]. This makes use of the differing Roche limits for silicate and icy bodies, where the Roche limit is the minimum separation between a moon and a planet beyond which the tidal forces from the planet exceed the moon's self-gravity, causing the moon to disintegrate. The Roche limit is given by the following equation [12]:

$$R_L \sim 2.456 R_p (\rho_p / \rho_M)^{1/3} \qquad (2)$$

Where $R_L$ = Roche limit, $R_p$ = planetary radius, $\rho_p$ = planetary density and $\rho_M$ = satellite density and the fluid approximation is used to more accurately account for the satellite's deformation. Assuming a typical ice density (930 kg m$^{-3}$) and silicate density (3,000 kg m$^{-3}$), substituting the appropriate values for Saturn pure ice moons could accrete at distances near the outer edge of the rings at 136,000 km from Saturn while silicate moons could form closer in at 90,000 km and accrete ice around themselves, filling their Hill sphere, before their interactions with the ring cause them to rapidly migrate outwards [12]. This mechanism is compatible with a wide range of ring ages from 4.5-2.5 Gyr hence permitting formation during the LHB or an early instability [8] and also preferentially removes dark material from the ring.

Cassini observations have also proved consequential to the number and positioning of moons in the early Saturn system. Cassini measured Titan's orbital migration to be 11.3 cm year$^{-1}$, a value far greater than expected which lent support to resonance locking, a model of tidal migration driven by the evolution of Saturn's interior [13]. At the time of an early instability resonance locking would indicate that Titan's semi-major axis was approximately $1/3^{rd}$ of its present day value of ~21 Saturn radii while the moon Rhea may also have been present: during the LHB Dione and Tethys would also have emerged from the rings while Titan would have migrated further to half its present day axis [13].

### Interactions Between Captured Bodies and Rings

Gravitational braking from moon capture is expected to perturb planetary ring systems: as energy is conserved the captured satellite must transfer orbital energy to the existing planetary system, which may include leaving detectable signatures on the rings. The distinct but related case of present-day interaction between rings and captured comets offers insights into this process.

Cassini observations made in August 2009, "when the Sun illuminated the rings from almost exactly edge-on," revealed regular radial variation of brightness across the C ring of Saturn attributed to a vertical corrugation in this ring; analysis of these images enabled calculation of the amplitude (2-20 m) and wavelength (30-80 km) of the corrugation [5]. These spiral patterns may be explained if the rings had been tilted (by $2 \times 10^{-8}$ to $3 \times 10^{-7}$ radians) with respect to Saturn and the "shear" (caused by inner ring particles orbiting faster than those further out) caused the spiral to become more tightly wound over time but nothing within the Saturn system seemed capable of causing this tilt in the required timescale [5].

The Jovian rings contributed towards the solution, beginning with 1996 Galileo observations showing a radial variation in brightness across the rings of Jupiter, which on closer inspection was attributed to a pair of vertical "ripple patterns" in the rings with differing wavelengths [6]. Further imaging in 2000 and 2007 supported the predicted relationship between wavelength and age, enabling the authors to link the formation of the longer wavelength ripple with the break-up of the comet Shoemaker-Levy 9, an interplanetary body captured and broken up by Jupiter in July 1994 [6]. While solid fragments struck the planet, the dust grains from the broken comet instead followed a trajectory where they were "deflected" into the ring under solar radiation pressure; these particles collectively possessed the required mass and velocity to change the ring's angular momentum and produce the observed tilt [6].

The Saturn corrugation was also attributed to an interplanetary origin potentially produced by a cloud of cometary debris with mass of $10^{11}$ to $10^{13}$ kg tilting the ring in 1983, at a time when Saturn's position made observation difficult [5]. The spiral patterns may hence be used to constrain parameters such as the interaction date, impactor trajectory and mass even when the interaction is not directly observed [6].

### SECTION 3: METHODOLOGY

#### The Nice Model

The methodology for the capture of Phoebe begins with a simulation of the Nice model to obtain data on interactions between Saturn and KBOs from the disk. This project utilizes the specialist astrophysical code REBOUND, chosen for its versatility and suitability for modelling the various stages of capture and evolution.

This project follows the currently accepted consensus model based on the 5GP which proved most successful: in this model the giant planets in the gas disk, which have initially circular coplanar orbits, migrate into a resonant chain with each pair of adjacent planets in a 3:2 resonance [7]. The orbits of the Sun and planets were set up in REBOUND's accurate adaptive default integrator IAS15 [14]. In order to migrate the giant planets into this chain in a manner consistent with prior work in the field a "fictitious force" was applied to simulate the gas disk [7]. Jupiter began 0.5 AU outside of a pre-instability orbit at 5.6 AU and each subsequent planet was 1% out of resonance with the inner planet so that the inner resonance forms during migration before the outer resonance: the planets are then integrated for ~1 Myr with a 6.0 day timestep to check that they are in resonance [8]. In order to implement the fictitious force the additional effects package REBOUNDx was required, specifically by applying modify_orbits_forces to the semi-major axes and eccentricities of the planets. This process excited the eccentricity of the innermost ice giant to 0.083, a value consistent with prior work [7]. This package was then disabled for subsequent stages as by this time the gas disk would have dispersed and it would no longer act to stabilize planetary orbits [7].

As shown in Figure 2, a disk of 10,000 planetesimals with a total mass 20 $M_\oplus$ was added between ~1.5 AU outside of Neptune's orbit and 30 AU [8]. This number of particles contributes towards a realistic and high resolution primordial disk [8] while the semi-major axes of the KBOs were chosen for a surface density profile of $1/r$ consistent with prior work [7] [8] and their eccentricities and inclinations are from Rayleigh distributions [8]. These particles were set as "semi-active", so that they would not interact with each other short of a direct collision.

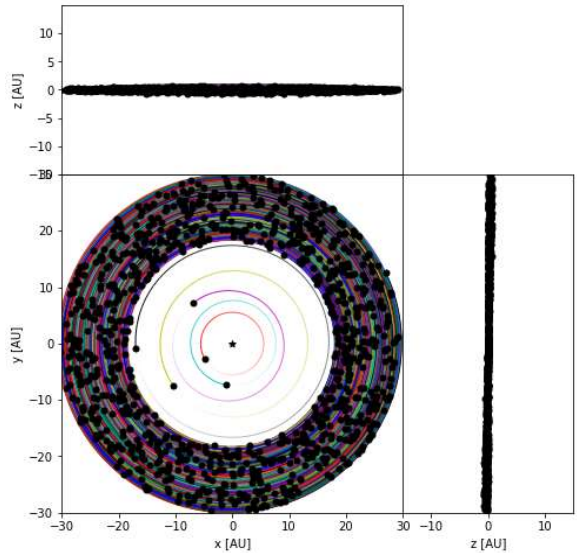

**Figure 2: The Outer Solar System Pre-Instability**

The instability (Figure 3) is artificially triggered by breaking the resonance between Saturn and the innermost ice giant; this is done by directly altering the mean anomaly of this ice giant by ~90° as it is thought that the asteroid belt was influenced by an interaction between this planet and either Jupiter or Saturn [8]. The system is then integrated for 100 Myr with a 50.0 day timestep [8]. Built-in collision detection records instances of interactions between planets and planetesimals. This integration was performed in REBOUND's MERCURIUS integrator [15], a hybrid integrator comparable to that used in previous Nice model studies [8] which offers improved speed relative to the compatible IAS15 integrator.

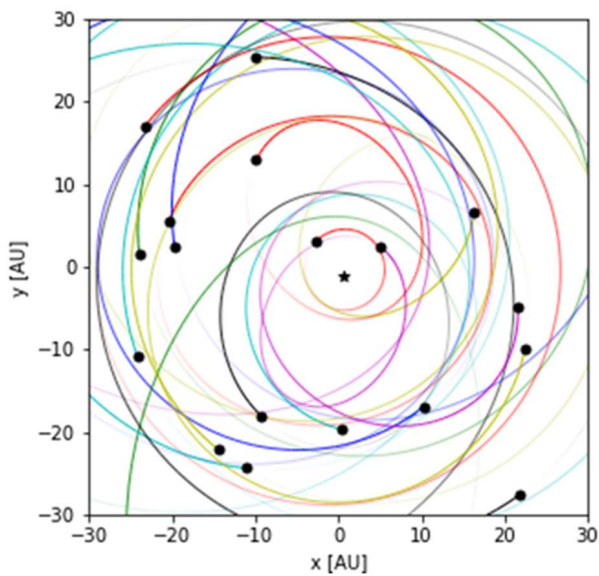

**Figure 3: An Instability Test With 20 Comets**

The recorded interactions are between Saturn and individual KBOs, hence for three-body capture they must be "re-enacted" replacing the original KBO with a binary object: to obtain the highest capture probability equal mass binary components are used while the semi-major axis of the circular orbit is set to 37,146 km [2].

As integrations only infrequently result in instabilities a continuing accumulation of runs is needed to identify those with higher success.

*Irregular Moon Capture*

From data on the tidal disruption of binary planetesimals and the capture or ejection of their components it is possible to find the orbit of a captured moon immediately after capture [2].

This modelling follows the assumption that Saturn's rings formed from a partially differentiated progenitor five times the mass of Saturn's moon Rhea with ice and silicates in a 4:1 ratio [11]. This initial mass of 1.15325 x $10^{22}$ kg is approximately 308 times the mass of Mimas hence this ring mass would evolve to the observed mass over the life of the Solar System [4]. Regular moon positions will be based on a resonance locking model with the primordial moon Titan present during an early instability while Rhea will also have emerged from Saturn's ring system [13].

A successful simulation would then evolve the post capture orbit of Phoebe towards its present day orbit [1]. As Phoebe's orbit has not been circularized by tidal forces alternative evolution methods are required, such as the effects of Phoebe's extensive collisional history [1] and via solar perturbations [2].

As REBOUND tracks the total energy of the modelled Saturn system, subtracting the orbital energies of the moons would show how the gravitational braking affects the global angular momentum of the ring and may also increase the angular momentum of Saturn itself. The angular momentum change for a ring is proportional to the mass and velocity of the impacting material [5] [6], enabling determination of the feasibility of tilting a ring in this scenario. The mass required to tilt Saturn's rings in the present day (~$10^{13}$ kg) [5] is a relatively small fraction of Phoebe's 8.29 x $10^{18}$ kg mass [1]. A more massive ring would have greater self-gravity and may hence dampen perturbations, however such rings may also be more gravitationally unstable. If this gravitational braking pulls material out of the Roche limit this could aid in the formation of regular moons. As a ring corrugation once formed evolves over time [8] it should be possible to discern whether the effects on the ring are lasting and visible in the present day and if not now when moon capture could be inferred from observation.

**SECTION 4: SUMMARY**

The aim of this project is to explore for the first time interactions between an irregular moon and planetary rings during capture.

This report has placed the capture of irregular moons in the context of the current Nice model and described a three-body capture mechanism of binary dissociation. This mechanism is applied to the specific representative irregular moon Phoebe to create the methodology for a model of this moon's history whereby the moon began as one component of a binary planetesimal originating in the outer Solar system which underwent an interaction with Saturn during the Nice model instability and dissociated at a distance of 299 Saturn radii from Saturn, enabling Phoebe's capture and subsequent evolution via interaction with the Saturn system.

The structure of the primordial Saturn system differs from the current day system: the ancient ring would be significantly more massive and is expected to be the source of Saturn's regular moons up to Titan: these regular moons are themselves now thought to have migrated outwards on rapid timescales and were formed at a fraction of their observed semi-major axes. The gravitational braking and orbital evolution of a Phoebe mass moon could feasibly tilt this ring inducing a long-lasting spiral corrugation.

This model may prove broadly applicable to other large irregular moons including exomoons.

*Acknowledgments*

The author would like to thank Matthew Clement for his advice on the Nice model instability and Hanno Rein for his support with REBOUND.